# Observation of Stable Bimeron Transport Driven by Spoof Surface Acoustic Waves on Chiral Metastructures

Huaijin Ma[1†], Te Liu[1], Jiachen Sheng[1], Kaiyan Cao[1*], Jinpeng Yang[1*], Jian Wang[1*]

*1. College of Physical Science and Technology, Yangzhou University, Yangzhou 225002, China*

† First author: Huaijin Ma (dx120250065@stu.yzu.edu.cn)

* Corresponding author: Kaiyan Cao (kycao@yzu.edu.cn), Jinpeng Yang (yangjp@yzu.edu.cn), Jian Wang (wangjian@yzu.edu.cn)

**Abstract:**

Topological quasiparticles, such as merons and bimerons, are characterized by non-trivial textures that exhibit remarkably robust transport against deformation, offering significant potential for information processing. While these phenomena have been explored in various systems, acoustic realizations remain challenging. Here, we report that acoustic meron topological textures were successfully realized using designed Archimedean-like square spiral metastructures via the excitation of spoof surface acoustic waves (SSAWs). By applying mirror-symmetric combinatorial operations to the unit structures, we further construct composite chiral metastructures that enable both one-dimensional and two-dimensional stable transport of acoustic bimerons. It is further revealed that bimeron transport originates from the locked opposite phase differences of SSAWs, induced by the handedness of the cavity resonant modes. The intrinsic robustness of the meron textures against structural defects is confirmed through the calculation of their topological charge. Our findings establish stable acoustic bimeron transport as a topologically resilient foundation for future acoustic information processing and storage technologies.

**Keywords:** Acoustics; Topological Bimeron; Transport; Chiral Metastructure

Topological textures[1, 2] serve as fundamental objects in the study of topological states of matter, capable of hosting or inducing a variety of exotic quasiparticles. Their non-trivial topological characteristics provide robust topological protections, endowing these quasiparticles for applications[1, 2] in data storage and logical operations. Since Tony Skyrme first proposed[3] skyrmions with integer topological charge in the sigma model in 1926, topological structures such as skyrmion[2, 3] and meron[1, 4] have been discovered across diverse fields including particle physics[5, 6], magnetic materials[7, 8] and quantum systems[9-11]. These stringent magnetic[7, 8] and optical[2, 12] experimental constraints significantly impede the deeper exploration of topological states of matter. Therefore, identifying readily excitable vector fields and developing methods for the macroscopic construction or observation of their topological textures are pivotal for advancing studies of topological phenomena and quasiparticles.

Acoustic systems provide a unique macroscopic platform for exploring topological phenomena of skyrmions[13-15] and merons[16, 17] using velocity vector fields. For instance, a hexagonally symmetric acoustic metasurface coupled with a six-speaker array[13] successfully generated skyrmion textures. Moreover, the bimeron texture—comprising a meron and an anti-meron with opposite polarities[1]—remains stable propagation[18] in curved geometries, offering a compelling alternative to skyrmions for next-generation spintronic devices. However, its exploration in acoustics remains limited, even though stationary meron and anti-meron textures have been demonstrated using standing waves[16] generated by a four-speaker array or by Mie-resonant metastructures[17] in isolated resonant cavities. Their transport and the underlying manipulation dynamics remain largely unexplored, and realizing bimeron transport together with a deeper understanding of its dynamics is essential for leveraging these quasiparticles in topological information storage and transmission.

In this work, we design Archimedean-like square spiral metastructures (ASSMs) that experimentally generate acoustic meron topological textures on metasurfaces through the excitation of spoof surface acoustic waves (SSAWs). We further demonstrate that composite chiral metastructures, formed via mirror-symmetric combinatorial operations, achieve stable one- and two-dimensional topological bimeron transport. We reveal that the bimeron transport originates from the locked, opposite phase differences of SSAWs dictated by the handedness of the cavity resonant modes. The intrinsic robustness of the meron textures against structural defects is validated through the calculations of their topological charge. Our findings establish stable acoustic bimeron transport as a topologically resilient foundation for future acoustic information processing and storage technologies.

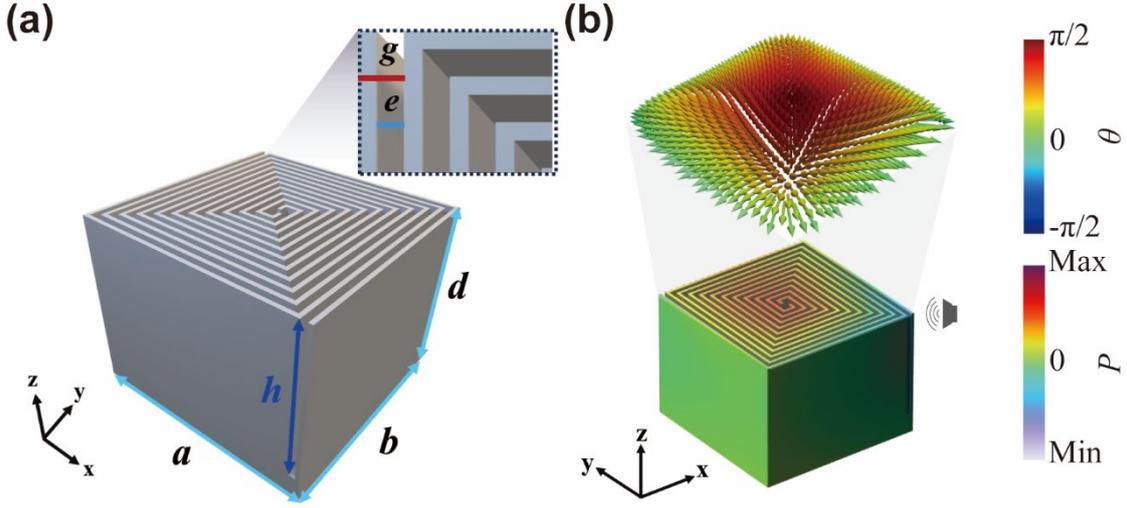

**Fig. 1 Schematic of the metastructure and meron mode. (a) Archimedean-like square spiral metastructure (ASSM). (b) Simulated meron textures: sound velocity vector field represented by conical arrows (Top) and sound pressure field visualized via color mapping (Bottom).** $\theta$ **denotes the angle between the velocity vector and the x–y plane.**

Acoustic metamaterials with Archimedean spiral or similar coiled geometries can efficiently manipulate wavefronts, enabling precise control[19-21] of sound propagation through refracted or reflected waves with suppressed amplitude, and facilitating sound localization via guided acoustic waves. Extensive studies show that spiral metastructures exhibit pronounced acoustic localization, facilitating the robust formation of topological textures in velocity vector fields[14, 15, 22]. Furthermore, accounting for the coupling between structural units, square geometries are more suitable than circular ones for constructing supercells via mirror symmetry, translational symmetry, and related operations. In addition, the double-open-ended spiral configuration promotes efficient transport between adjacent units[15]. To enhance this effect and enable inter-unit coupling, we designed an ASSMs as the acoustic building block for observing meron topological textures, as shown in Fig. 1 (a). The key to achieving topological texture through acoustic wave-metastructure coupling lies in the utilization and manipulation of spoof surface acoustic waves (SSAWs), which originate from[23] surface waves trapped by the metasurface and decay exponentially above the surface with the following SSAW wave vector $k_{ssaw}$,

$$k_{ssaw} = k_0\sqrt{1+\left(\frac{e}{g}\right)^2 \tan^2(k_0 h)}. \quad (1)$$

Here the wave vector in air is given by $k_0 = \dfrac{\omega}{c_s}$. The speed of sound $c_s$ = 343 m/s, and the angular frequency $\omega$ = $2\pi f$. The parameters $e$ and $g$ in Eq. (1) denote the channel width and wall thickness. To describe the stable acoustic eigenmodes on the metasurface, the sound pressure field $p$ and the velocity field $v_x, v_y, v_z$ can approximately be expressed as follows, respectively (see online Supplementary Material Section S I),

$$p = A\cos(k_x^s x)\cos(k_y^s y)\exp(-ik_z^s z), \tag{2}$$

$$v_{x,y,z} \propto \begin{bmatrix} k_x^s \sin(k_x^s x)\cos(k_y^s y) \\ k_y^s \cos(k_x^s x)\sin(k_y^s y) \\ \tau \cos(k_x^s x)\cos(k_y^s y) \end{bmatrix}, \tag{3}$$

where $\tau$ is the decay constant along the z-direction. Based on the SSAW wave vector $k_{ssaw}$, the relationship $k_{ssaw} = \sqrt{(k_x^s)^2 + (k_y^s)^2}$ holds for Eqs. (2) and (3). To achieve supercell operations of the acoustic unit along the x- and y- dimensions, the fixed boundary conditions of the structural units can be defined by the number of units $n_x$ and $n_y$ as follows,

$$\begin{cases} k_x^s L_x = n_x \pi, n_x = 1, 2, 3... \\ k_y^s L_y = n_y \pi, n_y = 1, 2, 3... \end{cases}, \tag{4}$$

where $n_x$ and $n_y$ denote the number of units along the x- and y-directions, respectively. The geometrical parameters of the metastructure are defined by its length $L_x$ and width $L_y$.

To verify the theoretical predictions, numerical simulations were performed using the Pressure Acoustics, Frequency Domain module in COMSOL Multiphysics. The geometry parameters of the ASSMs shown in Fig. 1(a) are $a = b = 9.2$ cm, $d = 7$ cm, and $h = 6$ cm. The channel width $e = 0.25$ cm, and the spiral pitch $g = 0.4$ cm. The resonant cavity was treated as an ideal rigid structure, while the channel and surrounding domain were modeled as air. First, we analyze the excitation frequency based on the structural parameter, the SSAW wave vector have relation to $k_{ssaw} = \sqrt{2}k_{x,y}^s$ when the length $a$ and width $b$ are equal. When considering the first-order eigenmode of a single unit, we have $L_x = L_y = 9.2$ cm and $n_x = n_y = 1$. Based on these parameters and Eqs. (1) and (4), the excitation frequency can be derived as $f = 1157$ Hz. The simulation results, shown in Fig. 1(b), reveal that the sound pressure distribution in the x-y plane reaches its maximum at the structural center (0,0) and gradually decreases to zero toward the edges. In the sound velocity vector field, the π/2 phase difference between the z-component and the x- and y-components causes the velocity vectors to be vertically oriented at the center. As they expand outward, the vectors undergo a π/2 rotational transition and ultimately become horizontally oriented near the edges. These observations indicate that the characteristics are in good agreement with the theoretical predictions based on Eqs. (2) and (3). The topological characteristics of the meron textures[24] are thus determined by combining the sound pressure and velocity vector fields. In contrast to topological textures generated through acoustic interference, the eigenmode-based excitation employed here requires only frequency-specific driving and remains robust against variations in parameters such as the sound-source position.

As a fundamental topological invariant, the topological charge $S$ provides a key quantitative measure for characterizing vector-field–based topological textures and is defined as follows [25]:

$$S = \frac{1}{4\pi} \iint \vec{n} \cdot \left( \frac{\partial \vec{n}}{\partial x} \times \frac{\partial \vec{n}}{\partial y} \right) dxdy, \quad (5)$$

Here the normalized vector field is defined as $\vec{n} = \frac{\mathrm{Re}(\vec{v})}{|\mathrm{Re}(\vec{v})|}$, where $\vec{v}$ represents the sound velocity vector. A nonzero topological charge $S$ reflects the presence of a nontrivial topological structure. The meron texture is characterized by a $\pm\pi/2$ flip in the velocity vector field, which maps the two-dimensional plane onto a three-dimensional hemisphere, resulting in a topological charge of $S = \pm 0.5$[1, 24]. The calculated topological charge $S$ is 0.49 in Fig. 2(b), confirming the meron nature of the texture.

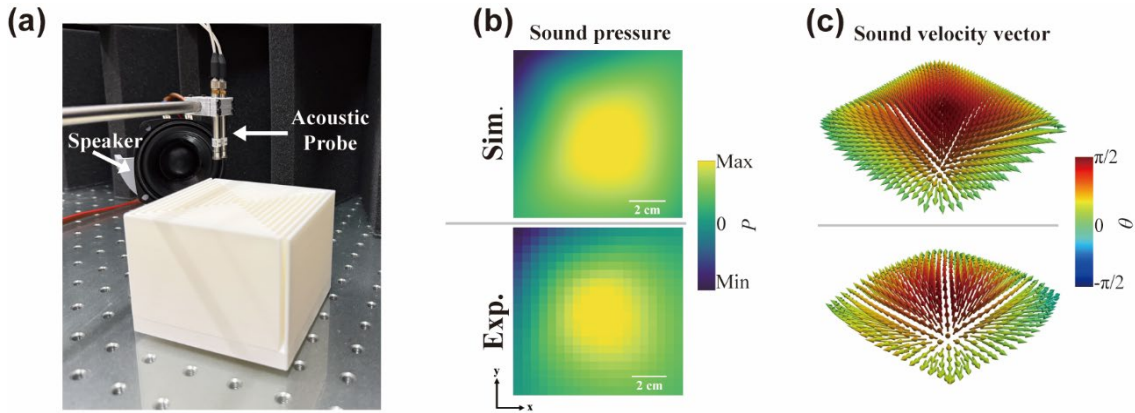

**Fig. 2 Comparison between experimental measurements and simulations of the ASSMs. (a) Photographs of the fabricated sample and experimental setup. (b) and (c) Simulated and measured meron mode under single acoustic source at frequency $f$ = 1157 Hz: (b) The sound pressure field on the metasurface, (c) The distribution of sound velocity vectors represented by conical arrows. Here $\theta$ is the angle between the velocity vector and the x–y plane.**

To experimentally observe the meron texture, the ASSM shown in Fig. 2(a) was fabricated via 3D printing using the same geometric parameters as those used in the simulations. Excitation was realized by placing a single acoustic source at the outer corner, as shown in Fig. 2(a), serving as a monopole emitter coupled to the metastructure to generate acoustic radiation at $f = 1157 \mathrm{Hz}$. The topological configuration was obtained by extracting the sound pressure and velocity vector fields in the x-y plane. To verify the reliability of the structural design, numerical simulations were conducted using the Pressure Acoustics, Frequency Domain module in COMSOL Multiphysics. The simulation results are displayed in the top panels of Figs. 2(b) and 2(c). It can be found that the peak sound pressure of the SSAW in Fig. 2(b) gradually decreases from the center of the structure toward the edges, approaching zero at the boundaries. The conical arrows in Fig. 2(c) represent the distribution of the sound velocity vector field above the structure, where the color map indicates the flip angle $\theta$ of the vectors in the x-y plane. Obviously, the direction of the sound velocity field continuously rotates from $\pi/2$ to 0, exhibiting the characteristic features[2] of a meron texture.

In the experiment, a loudspeaker was placed 7 cm from the sample at the outer corner. Acoustic signals were produced using a signal generator (33500B, Keysight) and amplified by a power amplifier (PX3, YAMAHA) to drive the loudspeaker. Sound pressure data of SSAWs were acquired using an acoustic probe (378C01, PCB Piezotronics). The sound velocity vector was subsequently calculated based on the relation $\vec{v} = -\frac{1}{\rho_0}\int \nabla p \, dt$. It can be seen that the measured SSAW sound pressure distributions, presented in Fig. 2(b), is in good agreement with the simulations. The experimental sound velocity vector distributions of SSAWs, shown in Fig. 2(c), are consistent with the simulation results. Clearly, both measured sound pressures and velocity vectors confirm the topological nature of the acoustic meron texture.

Next, we investigate the one-dimensional and two-dimensional propagation of acoustic bimerons. A bimeron comprises two coupled merons with opposite polarities[4, 26], whose distinctive vector field configuration offers substantial advantages for information transmission[17, 18]. Realization of bimerons requires the design of suitably coupled ASSMs. Based on the previously described single ASSM, we construct a pair of composite metastructures, denoted '0' and '1', by applying a mirror operation across the dashed line, as shown in Fig. 3(a). Composite chiral metastructures were thus constructed via a mirror operation to achieve cavity[21, 27] resonant modes with distinct chiral properties. In the '0' unit, sound waves traverse a clockwise spiral (CS) and exit via a counterclockwise spiral (CCS), defining the CS–CCS vortex as a left-handed (LH) chiral property. In the '1' unit, sound waves propagate along CCS–CS in contrast, exhibiting right-handed chirality (RH). As shown in Fig. 3(a), simulations confirm that both the sound pressure field (middle) and the sound velocity vector field (top) on the chiral metastructure exhibit bimeron topological textures. The sound pressure consists of positive and negative poles, while the acoustic bimeron is characterized by a 360° rotation of the sound velocity vector field along the z-axis. The '0' unit gives rise to merons, while the '1' unit generates anti-merons[2, 24, 28].

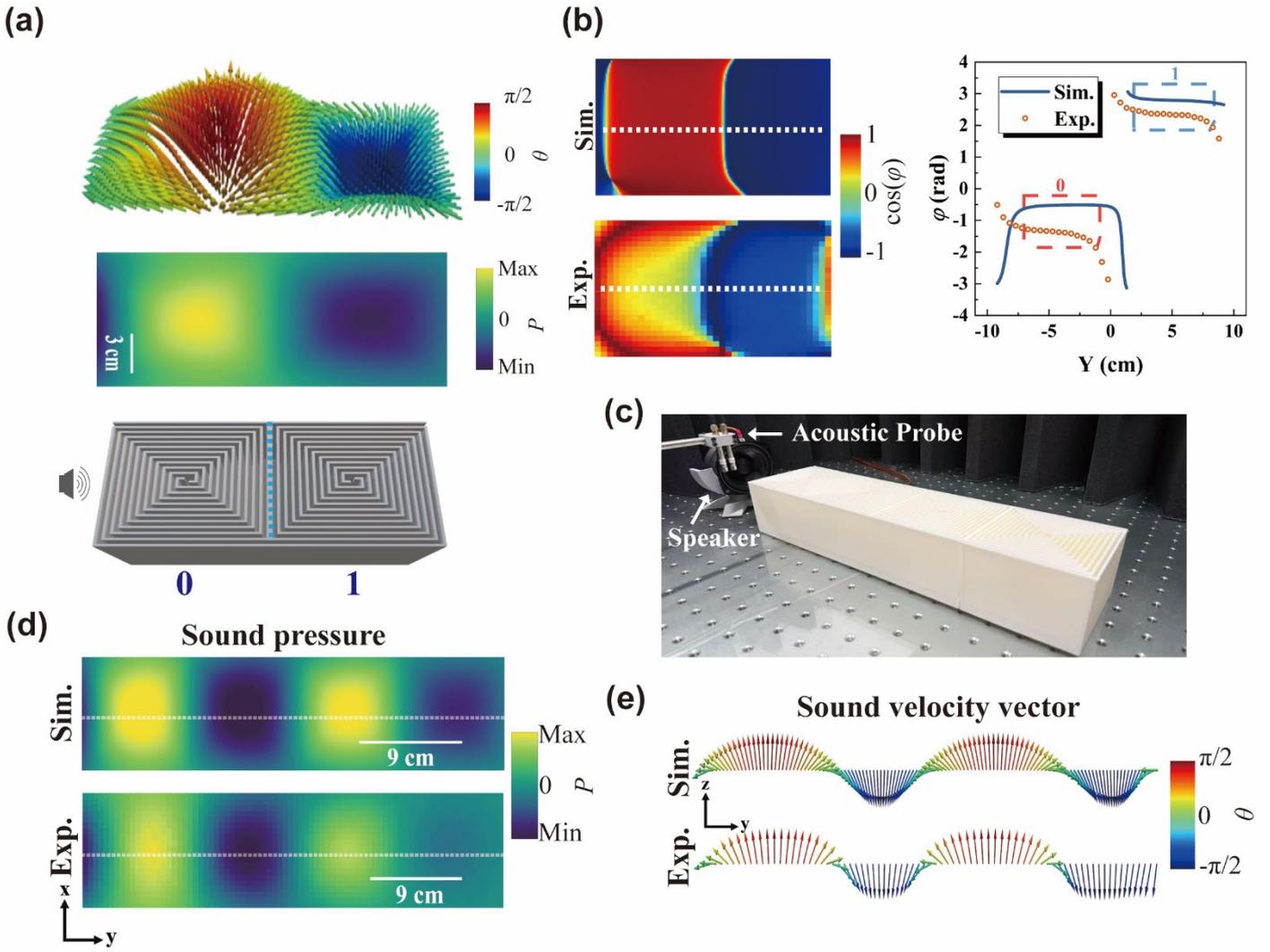

**Fig. 3 Observation of one-dimensional bimeron transport.** (a) Schematic of the composite metastructure (Bottom) and the simulated bimeron textures of SSAWs, showing the sound pressure field (Middle) and the corresponding velocity vector field (Top). (b) Simulated and measured phases of the SSAW field. (c) Experimental setup and fabricated samples consisting of four units. (d) Comparison of simulated and measured sound pressure fields for one-dimensional bimerons. (e) Simulated and measured distributions of sound velocity vectors along the dashed lines in Fig. (d).

To clarify the mechanism of bimeron generation by chiral metastructures, we present the simulated and experimental phase $\cos\varphi$ of the metasurface acoustic wave in the left panel of Fig. 3(b). As shown in the figure, the SSAW phases on the chiral units are clearly locked to two distinct states, as represented by two separate isosurfaces. The phase $\varphi$ profiles along the y-axis, indicated by the central dashed white line, are further plotted for both simulations and experiments in the right panel of Fig. 3(b). We observe that the phase difference of the SSAWs between the '0' and '1' units is approximately $\pi$; in other words, the SSAWs exhibit opposite phases on the two chiral metasurface units. These two opposite phases lead to the formation of merons and anti-merons, which possess opposite polarities[4, 26], respectively. We attribute the opposite phase difference of the SSAWs to the distinct chiral properties (LH and RH) of the cavity resonant modes in the chiral metastructure units. In the context of SSAW propagation, acoustic metamaterials incorporating chiral Archimedean spirals can effectively tailor[21] the wavefront, enabling precise control over the propagation characteristics.

In the experiment, the chiral metastructures are arranged in a linear chain to study one-dimensional bimeron propagation, as shown in Fig. 3(c). To achieve acoustic excitation, a single sound source with $f = 1157$ Hz is positioned on the left side of the chain. Figs. 3(d) and 3(e) illustrate the sound pressure distribution and the sound velocity vector field, respectively, obtained through simulations and experiments. Figure 3(e) shows sound velocity vectors calculated from the sound pressure fields along the dashed central line in Fig. 3(d). The sound pressure and velocity vector distributions clearly exhibit the characteristic features[4, 28] of bimeron structures. Together, the simulation and experimental results confirm the successful one-dimensional propagation of bimeron quasiparticles excited by a single sound source along the chain of chiral metastructures.

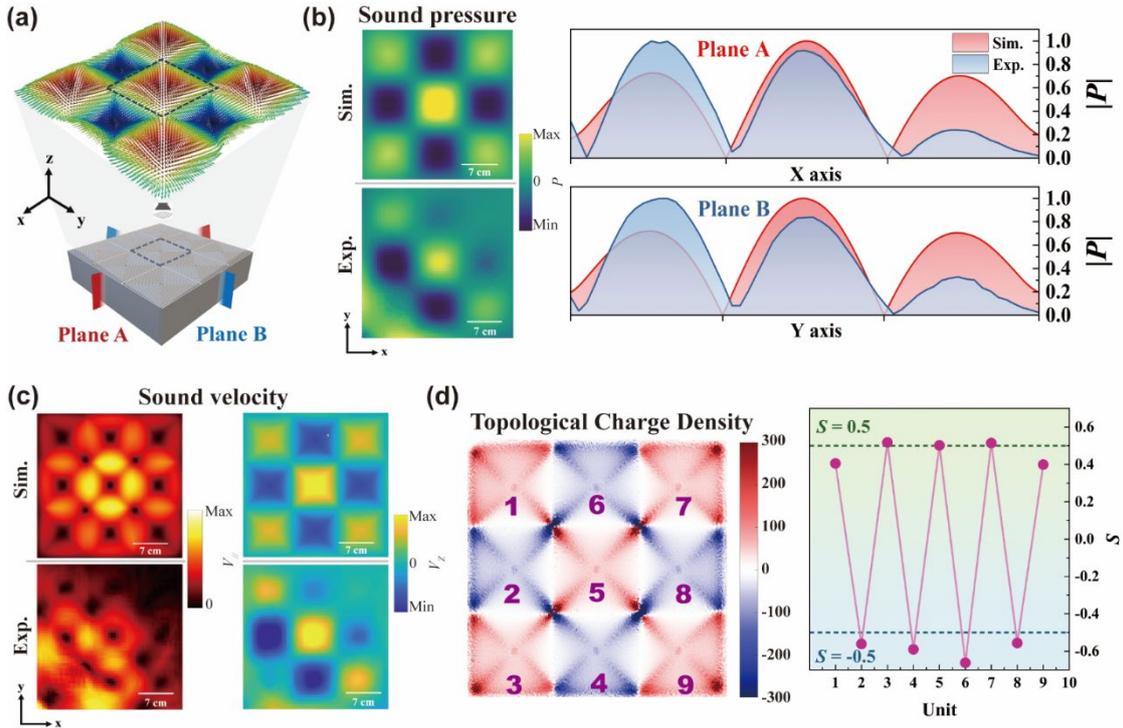

**Fig. 4 Observation of two-dimensional propagation of bimerons. (a)** Schematic of the fabricated sample (Bottom) and the simulated sound velocity vector distribution (Top). **(b)** Sound pressure fields on the metasurface (Left). The profiles of sound pressure fields (Right) along Planes A and B, respectively. **(c)** The distribution of sound velocity vector $v_{//} = \sqrt{v_x^2 + v_y^2}$ in the plane (Left) and $v_z$ along the z-direction (Right), respectively. **(d)** Topological charge densities and values for each meron and anti-meron texture.

Furthermore, the propagation of two-dimensional bimerons is achieved within a chiral metastructure array. Design details are provided in the Supplementary Material Section S II, with the structural schematic shown in Fig. 4(a). The upper panel of Fig. 4(a) presents the simulated sound velocity vectors under single-source excitation at 1157 Hz, where the bimeron textures are clearly discernible. This two-dimensional bimeron propagation displays a morphology consistent with meron lattices previously reported[28, 29] in optics and magnetism. Fig. 4(b) shows the simulated and experimental sound pressure distributions for this two-dimensional metastructure array. The normalized sound pressure $|P|$ is also plotted along Plane A and Plane B in Fig. 4(b), respectively. The distribution of sound velocity vector $v_{//} = \sqrt{v_x^2 + v_y^2}$ and $v_z$ along the z-direction from simulations and experiments are presented in Fig. 4(c). The figures demonstrate that the experimental results are in good agreement with the simulations, thereby confirming the two-dimensional bimeron propagation. Furthermore, the effects of varying sound-source positions are examined in detail (see online Supplementary Materials Section S II), demonstrating the topological robustness against sound source position variations.

A bimeron consists of one meron and one anti-meron, each carrying a topological charge $S$ [1] of ±0.5. To investigate the topological nature of the acoustic bimeron textures, we further computed the topological charge density[13] $\rho_S = \vec{n} \cdot \left( \frac{\partial \vec{n}}{\partial x} \times \frac{\partial \vec{n}}{\partial y} \right)$ and the corresponding topological charge $S$, shown in Fig. 4(d). The figure shows that the topological charge $S$ in each odd texture is approximately +0.5, whereas $S$ in each even texture is about −0.5, corresponding to the meron and anti-meron, respectively. These findings offer compelling topological evidence for the formation of bimeron quasiparticles.

To further validate the robustness of the acoustic meron topological texture, we designed the following experiment. As illustrated in Fig. 5(a), we introduced vacancy (V) and filling (F) types of defects into the metastructure. The simulated and measured distributions of the sound pressure field and sound velocity vector field for different defect types are presented in Figs. 5(b) and 5(c), respectively. Both experimental and simulated results show good agreement. No significant distortions are observed in the sound pressure or velocity vector distributions at the defect sites, indicating that topological protection is preserved. Moreover, the topological charge $S$ can serve as an effective metric[13] for assessing the robustness of topological structures against defects. Topological charge $S$ as a function of vacancy (V) and filling (F) defects are shown in Fig. 5(d). All calculated $S$ values are close to 0.5, indicating that the meron texture maintains stable topological robustness against defects.

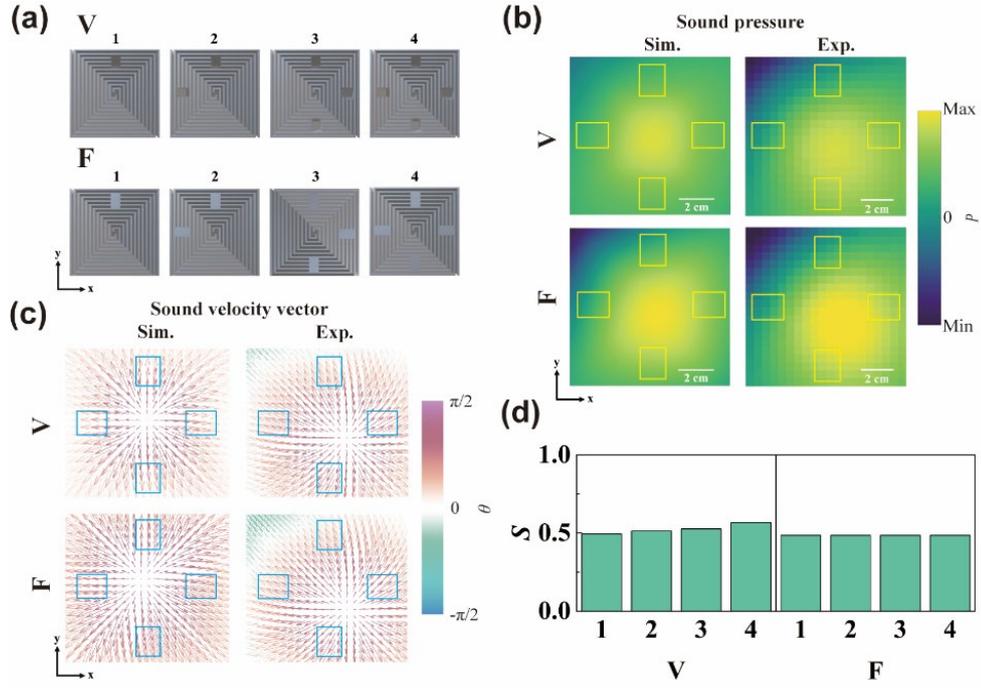

**Fig. 5 Robustness verification of meron topological textures.** (a) Schematic illustration of two types of defects: vacancy-type (V) and filling-type (F). (b) and (c) Simulated and measured distributions of the sound pressure field and sound velocity vector field for different defect types. (d) Topological charge $S$ as a function of vacancy (V) and filling (F) defects.

In summary, we successfully generated acoustic meron topological textures on metasurfaces via spoof surface acoustic waves, with their topology quantified by the associated topological charge. The SSAWs serve as a robust medium for observing and manipulating topological merons. In the composite chiral metastructures, cavity resonant modes with distinct handedness give rise to merons and anti-merons with opposite polarities. Consequently, stable bimeron transport is achieved in both one-dimensional and two-dimensional configurations, enabled by the locked opposite phase on each chiral metasurface unit. The robustness of the meron texture was further confirmed through the calculation of their topological charge. Our work not only paves the way for stable topological bimeron transport in acoustic systems but also lays the foundation for future applications in acoustic information storage and logic operations.

## Acknowledgements

J. Wang acknowledges the support from National Natural Science Foundation of China (NSFC) under Grants No. 11875047. K.Y. Cao was supported by the Jiangsu Funding Program for fundamental research under Grant No. BK20250886 and the Foundation of Yangzhou University under Grant No. 137013633. J.P. Yang acknowledges the support from National Natural Science Foundation of China (NSFC) under Grant No. 62375234.

**Data availability**

The data generated during and/or analyzed in this article are available from the corresponding author on reasonable request.

**Conflict of Interest**

The authors have no competing interests to declare that are relevant to the content of this article.

# Supplementary Material

**Section I** Therectical derivation of **Eq.**(2), **Eq.**(3) and **Eq.**(4)

Consider a square unit cell with side length *a* and height *h*, with the origin of the coordinate system positioned the center of the upper surface. Using the parameters of the metastructure outlined in the text, the corresponding dispersion relation is derived,

$$k_{ssaw} = k_0 \sqrt{1 + \left(\frac{e}{g}\right)^2 \tan^2(k_0 h)}. \tag{S1}$$

Here, $k_0 = \frac{\omega}{c_s}$, and $k_{ssaw} = \sqrt{(k_x^s)^2 + (k_y^s)^2}$ is defined as the in-plane SSAW vector in the region above the metastructure. For the out-of-plane wave vector, $k_z^s = \sqrt{(k_0)^2 - (k_{ssaw})^2}$. Since $k_0 < k_{ssaw}$, $k_z^s = -i\tau$ is purely imaginary, suggesting that the SSAWs decay exponentially along the z-direction. Considering the acoustic wave on the upper surface of the metastructure, it can be treated as an equivalent rectangular resonant cavity that satisfies the Helmholtz equation,

$$\frac{\partial^2 p}{\partial x^2} + \frac{\partial^2 p}{\partial y^2} + \frac{\partial^2 p}{\partial z^2} = \frac{1}{c_s^2} \frac{\partial^2 p}{\partial t^2}. \tag{S2}$$

Now, let the solution be:

$$p = p_a(x, y, z) \exp(i\omega t). \tag{S3}$$

Substituting into the equation yields:

$$\frac{\partial^2 p_a}{\partial x^2} + \frac{\partial^2 p_a}{\partial y^2} + \frac{\partial^2 p_a}{\partial z^2} = k_0^2 \frac{\partial^2 p_a}{\partial t^2}, \tag{S4}$$

By applying separation of variables to the equation, let:

$$p_a(x, y, z) = X(x) Y(y) Z(z), \tag{S5}$$

Three independent ordinary differential equations in their respective coordinates are obtained:

$$\begin{cases} \dfrac{\partial^2 X(x)}{\partial x^2} + k_x^{s2} X(x) = 0 \\ \dfrac{\partial^2 Y(y)}{\partial y^2} + k_y^{s2} Y(y) = 0 \\ \dfrac{\partial^2 Z(z)}{\partial z^2} + k_z^{s2} Z(z) = 0 \end{cases} \quad \text{(S6)}$$

On the upper surface of the metastructure, boundaries exist. For $X(x)$ and $Y(y)$, the following forms of solutions are adopted:

$$\begin{aligned} X(x) &= A_x \cos(k_x^s x) + B_x \sin(k_x^s x) \\ Y(y) &= A_y \cos(k_y^s y) + B_y \sin(k_y^s y) \end{aligned}, \quad \text{(S7)}$$

For the z-direction extending to infinity, a traveling wave solution is assumed:

$$Z(z) = A_z \exp(-ik_z^s z). \quad \text{(S8)}$$

At the edges of the structure, the acoustic pressure amplitude approaches zero. On the boundaries of the equivalent rectangular resonant cavity, a soft acoustic boundary condition is approximately applied, i.e., $p\big|_{x,y=\pm\frac{a}{2}} = 0$,

$$\begin{aligned} A_x &= 0, k_x^s L_x = n_x \pi, n_x = 1, 2, 3... \\ A_y &= 0, k_y^s L_y = n_y \pi, n_y = 1, 2, 3... \end{aligned}. \quad \text{(S9)}$$

Therefore, the sound pressure distribution on the surface of the metastructure is:

$$p_{n_x n_y} = A_{n_x n_y} \cos(k_x^s x) \cos(k_y^s y) \exp(-ik_z^s z) \exp(i\omega t). \quad \text{(S10)}$$

This indicates the presence of a stable acoustic wave mode on the surface of the metastructure. Its amplitude is determined by $A_{n_x n_y} \cos(k_x^s x) \cos(k_y^s y)$, while in the z-direction, the air medium exhibits an overall vibration with exponential decay.

From the linear acoustic Euler equation:

$$\rho_0 \frac{\partial v}{\partial t} = -\nabla p, \quad \text{(S11)}$$

$$v = -\frac{-i\nabla p}{\omega \rho_0}, \quad \text{(S12)}$$

Substituting the sound pressure expression yields the velocity components in the same plane:

$$v_{x,y,z} \propto \begin{bmatrix} k_x^s \sin(k_x^s x)\cos(k_y^s y) \\ k_y^s \cos(k_x^s x)\sin(k_y^s y) \\ \tau \cos(k_x^s x)\cos(k_y^s y) \end{bmatrix}. \tag{S13}$$

**Section II** Design of two-dimensional composite chiral metastructure and their robustness to the sound source

*Two-dimensional chiral metastructure.* Based on the two-unit composite metastructure constructed via "01" encoding, a fully interconnected 3×3 array was formed by applying mirror symmetry to the boundaries of adjacent units arranged in a "010101010" sequence, as shown in Fig. S1. This configuration can also be interpreted as a one-dimensional chain of nine units that has been folded twice to form the metastructure.

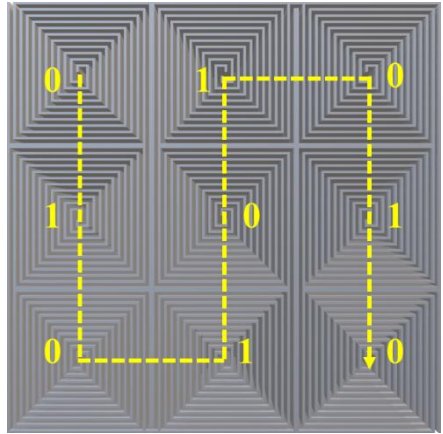

**Fig. S 1 Schematic diagram of the two-dimensional 3×3 array coupling design**

*Robustness to the sound source.* The excitation of meron textures is inherently dependent on the structural eigenmodes, such that the resulting two-dimensional coupled array is, in theory, independent of the position of the sound source within the acoustic-material interaction. This independence is contingent upon satisfying the excitation frequency of the eigenmode. Based on this principle, we devised two distinct sound source placement configurations, as illustrated in Fig. S2.

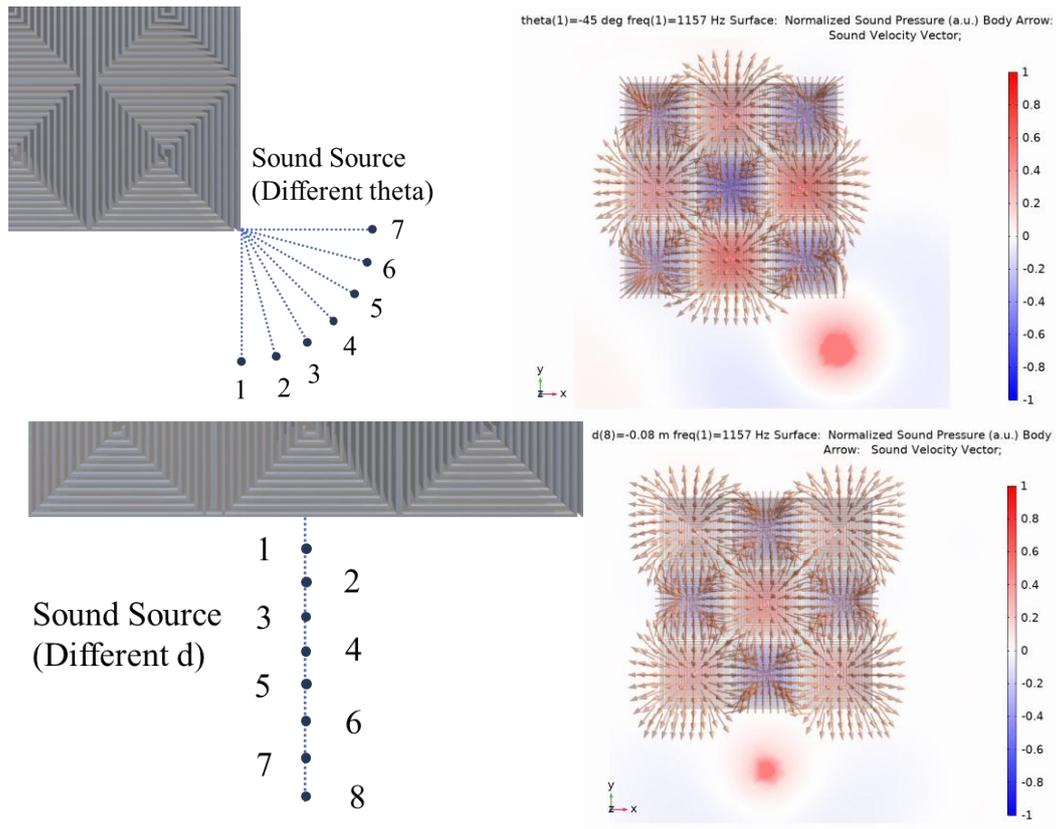

**Fig. S 2 Two sound source placement schemes**

Initially, for diagonal excitation, the sound source was positioned 7 cm from the sample at the outer corner, with seven distinct angular placements at 15° intervals. The accompanying animated sequence illustrates the sound velocity vector field and the corresponding surface sound pressure colormap for each excitation angle. In all angular configurations, the sound source effectively excites the meron texture lattice through coupling with the underlying structure.

Additionally, for edge excitation, the sound source was placed at the center of the edge, with eight distinct distances from the edge, each separated by 1 cm intervals. The accompanying animated sequence depicts the acoustic velocity vector field and the surface acoustic pressure colormap for varying excitation distances. Similarly, stable meron texture characteristics were observed under this excitation scheme as well.